\documentclass[aps,prl,twocolumn,superscriptaddress]{revtex4-2}
\usepackage{amsmath,amssymb,graphicx,xcolor}
\newcommand{\Trel}{\tau_{\rm relax}}
\usepackage{hyperref}
\begin{document}

\title{Time-delayed feedback turns Arrhenius escape logarithmic}

\author{Roy Podgaetsky}
\affiliation{School of Chemistry, Tel Aviv University, Tel Aviv 6997801, Israel}
\author{Vishwajeet Kumar}
\author{Arnab Pal}
\email{Corresponding author: arnabpal@imsc.res.in}
\affiliation{The Institute of Mathematical Sciences, C.I.T. Campus, Taramani, Chennai 600113, India}
\affiliation{Homi Bhabha National Institute, Anushakti Nagar, Mumbai 400094, India}
\author{Ohad Shpielberg}
\email{Corresponding author: ohads@sci.haifa.ac.il}
\affiliation{Haifa Research Center for Theoretical Physics and Astrophysics, University of Haifa, Haifa 3498838, Israel}

\begin{abstract}

Thermal escape is governed by the Arrhenius law, where the mean escape time scales exponentially with the barrier height. We show that the non-Markovianity induced by time-delayed feedback in the confining force removes this exponential scaling. Beyond a threshold set by the curvature of the minimum, the delay destabilizes the well, and the thermal noise seeds an instability that is subsequently amplified deterministically to the boundary leading to \textit{slingshot} escape trajectories. The escape time becomes logarithmic in the barrier, its fluctuations follow a Gumbel law, and an optimal delay enables escape faster than free diffusion. Our results propose time delay as a tunable and experimentally feasible control parameter for accelerating activated processes.

\end{abstract}

\maketitle

\emph{Introduction.}---Thermally activated escape over an energy
barrier is a cornerstone of statistical physics, underlying chemical
kinetics, nucleation, and biomolecular conformational change
\cite{arrhenius1889,kramers1940,hanggi1990}. Its signature is the
Arrhenius law: the mean escape time grows exponentially with the ratio
of the barrier height $\Delta U$ to the thermal energy,
$\langle\mathcal{T}\rangle\sim e^{\Delta U/k_BT}$. The law is
remarkably robust. It survives when the friction is made non-Markovian
through a memory kernel \cite{grote1980,kappler2018,berner2018}, in many-body and
interacting diffusive systems \cite{langer1969,kumar2024arrhenius,kumar2024,kumar2025inferring}, in glassy
landscapes with a hierarchy of barriers \cite{goldstein1969}, and in
active matter, where self-propulsion drives the bath weakly out of
equilibrium \cite{woillez2019,militaru2021}; in each case the activated,
exponential character persists, with only prefactors or effective
barriers modified.

Time delays arise generically wherever a force depends on the state of
the system at an earlier instant --- through finite signaling, sensing,
or processing times. Rather than entering the friction, as in
generalized-Langevin descriptions \cite{zwanzig2001,grote1980}, the
delay here acts in the conservative force: the force at time $t$ is set
by the particle's position a delay $\tau$ earlier. Delayed forces are
directly realizable and widespread: feedback-controlled optical traps
impose a delayed force on a colloidal particle with $\tau$ tunable in
software \cite{belldavies2023}; sensorial delay engineered
into active colloids controls their collective behaviour
\cite{mijalkov2016}; and gene-regulatory circuits \cite{bratsun2005}
realize the same structure in biology. A broad account of stochastic
systems with delay is given in \cite{loos2021book}. Such dynamics are
intrinsically non-Markovian and driven far from equilibrium, violating
Boltzmann statistics and the fluctuation--dissipation relation
\cite{Loos2017Force,Loos2019Heat,Rosinberg2015Stochastic,kopp2026inference}.

Delay has been studied mostly as a stabilizing agent. In time-delayed
feedback control \cite{pyragas1992,janson2004}, a delayed force is
applied alongside an instantaneous one and tuned to stabilize an
otherwise unstable state. 
Where the force is purely delayed, studies of stochastic escape have
stayed at small delay \cite{guillouzic1999,guillouzic2000} or characterized the noise-induced
switching statistics without destabilizing  the wells \cite{masoller2002noise,Masoller2003,Curtin}; where a delayed correction is added to
an otherwise stable potential, the coupling has been kept weak enough
that the wells never destabilize \cite{tsimring2001}. Delayed harmonic interactions in colloidal clusters reach the
same oscillatory instability, but the escape between configurations has
been treated only below threshold, as an effective Arrhenius process
over a delay-reduced barrier, where a stationary rate exists \cite{geiss2019brownian}. In all cases the delay renormalizes rates and correlation times, but not the mechanism: escape remains activated over an unmodified barrier. However,
at large delay the minimum itself loses stability, and the basic
questions of thermal activation reopen: What becomes of Arrhenius
escape? How do the barrier height and potential
shape enter the escape rate? Does activation remain a rare event driven
by large thermal fluctuations, or does this non-Markovianity reshape the
escape trajectories?  

\begin{figure}[t]
\centering
\includegraphics[width=0.95\linewidth]{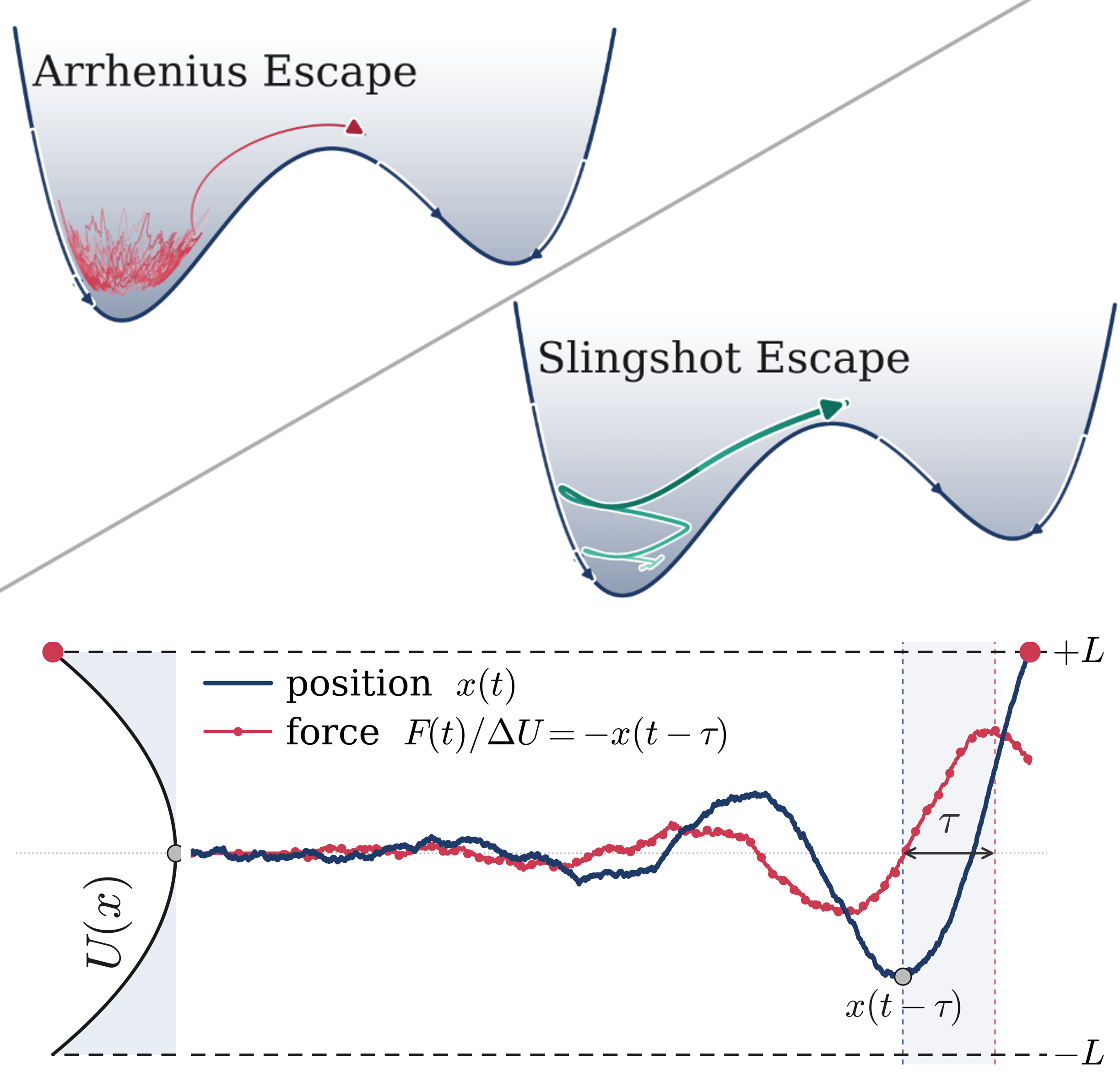}
\caption{\textbf{The slingshot mechanism.} 
\textbf{Top: Arrhenius escape}. Escape via a rare fluctuation (red) over the barrier, giving exponentially long escape times set by $\Delta U/D$.
\textbf{Middle: Slingshot escape}. Delayed feedback drives the particle over the barrier deterministically in a few exponentially increasing oscillations (green).
\textbf{Bottom: Slingshot trajectory.}
A typical trajectory of $x(t)$ (navy) with a feedback force $F/\Delta U = -x(t-\tau)$ (red). Dashed vertical lines mark $t-\tau$ and $t$. The delay $\tau$ induces a phase shift
between $x(t)$ and $F(t)$, driving the particle to the boundary $\pm L$.}
\label{fig:mech}
\end{figure}

To address these issues, in this \textit{Letter}, we consider a broad class of overdamped noisy systems in
which a particle diffuses under a delayed conservative force, and
characterize the dynamics by the dimensionless delay $s=\tau/\Trel$,
where $\Trel$ is the relaxation time at the stable minimum. Beyond a
critical delay $s_c=\pi/2$, the minimum loses stability through a Hopf
bifurcation: a typical thermal fluctuation is amplified by the delayed
feedback into a deterministic \textit{slingshot} trajectory, in stark
contrast to conventional thermal activation, which proceeds through
rare fluctuations (Fig.~\ref{fig:mech}). For a harmonic trap, we obtain a closed-form escape
law that is logarithmic in the barrier height, replacing the
exponential Arrhenius dependence, and derive the full escape-time
distribution. We then show that the coefficient of the logarithm is a
curvature invariant --- identical across a family of hardening
potentials. Therefore the law is not special to the harmonic case, and
the resulting speedup is exponential and controlled by the single
parameter $\tau$, establishing time-delayed feedback as a robust and tunable mechanism for accelerating activated processes.

\emph{Model.}---We consider an overdamped particle at position $x(t)$
in a confining potential $U$, with the force evaluated at the delayed position,
\begin{equation}
\gamma\,\dot x(t)=-U'\!\big(x(t-\tau)\big)+\sqrt{2\gamma^2 D}\,\eta(t),
\label{eq:model}
\end{equation}
where $\gamma$ is the friction, $\tau>0$ the delay in the force $F(x)=-U'(x)$, Gaussian
white noise $\eta$ of unit strength, and $D=k_BT/\gamma$ the diffusion
constant. We set $\gamma\equiv1$ throughout, so that $D$ coincides
with the thermal energy $k_BT$ and $\Delta U/D$ is the usual
dimensionless Arrhenius ratio.
Being non-Markovian, Eq.~\eqref{eq:model} requires an initial
history on $[-\tau,0]$; we take the particle to rest at the minimum,
$x(t)=0$ for $t\le0$, so that no force acts during the first interval of
length $\tau$ and the dynamics start from free diffusion. The choice of history sets the scale of the initial displacement but not
the escape mechanism, which is a property of the dynamics; the distribution shape is in fact independent of it \cite{SM}.  

Escape is defined as the first passage of the particle to $|x|=L$ and the corresponding random escape time is denoted by $\mathcal{T}$. In what follows, we first present an analytically solvable realization of delay-induced escape in a harmonic trap,
$U(x)=\Delta U\,(x/L)^2$, with barrier height $\Delta U$ at the absorbing boundaries. Measuring time in units of
$\Trel=\gamma/U''(0)$ and lengths in units of $L$ renders the problem
dimensionless, and leaves exactly two control parameters: the rescaled
delay $s=\tau/\Trel$, which governs the stability of the minimum, and
the barrier-to-noise ratio $\Delta U/D$, which measures how rare a
thermal excursion to the boundary is. In the weak-noise regime
$\Delta U/D\gg1$ the Arrhenius law is recovered at $\tau=0$. All results
below are reported in terms of the dimensionless escape time
$\langle\mathcal{T}\rangle/\Trel$ as a function of these two.

\begin{figure}[t]
\includegraphics[width=\linewidth]{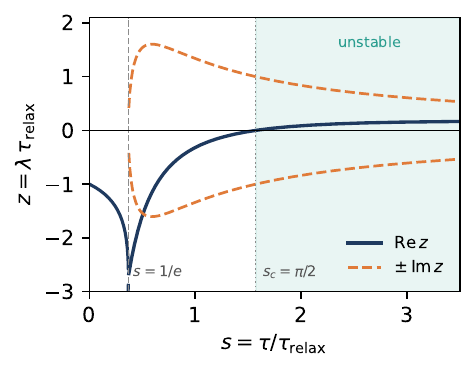}
\caption{\textbf{Structure of the leading root $z=\lambda\Trel$ of the characteristic equation
\eqref{eq:char}.} The root $z$ is complex in nature. For $s<1/e$ two real roots exist; they coalesce at
$z=-e$ and become a complex pair, whose real part (solid) crosses zero
at $s_c=\pi/2$. Beyond that (shaded) the minimum is an unstable focus
with growth rate $\mu$ and frequency $\omega$ (dashed).}
\label{fig:roots}
\end{figure}

\emph{Destabilization of the minimum.}--- 
Consider first the noiseless dynamics which gives
$\dot x=-x(t-\tau)/\Trel$, and the ansatz $x\propto e^{\lambda t}$
yields the transcendental characteristic equation 
\begin{equation}
z=-e^{-s z},\qquad z\equiv\lambda\Trel,
\label{eq:char}
\end{equation}
whose solutions are given by the branches of the Lambert $W$ function,
$z_k=W_k(-s)/s$ \cite{yi2010time}. The roots
$z_k$ depend on $s$ and are in general complex,
leading to a rich and qualitatively changing dynamical behavior as $s$ is varied. For $s<s_c$ every root has
negative real part and the minimum is stable; as $s$ increases, the
leading pair of roots crosses the imaginary axis. The crossing is a
Hopf bifurcation: at $s_c=\frac{\pi}{2}$  
a conjugate pair sits at $z=\pm i$, and for $s>s_c$ the minimum is a
linearly unstable focus --- an oscillatory instability with growth rate
$\mu=\mathrm{Re}\,\lambda>0$ and frequency $\omega=\mathrm{Im}\,\lambda$
(Fig.~\ref{fig:roots}). The delay introduces a phase lag: when $\tau$
is comparable to the oscillation period, the force reflects a position
the particle has already left, and negative feedback becomes positive
(Fig.~\ref{fig:mech}).

The threshold $s_c=\pi/2$ is universal as  it depends on the potential only through the local curvature
$U''(0)$ contained in $\Trel$; equivalently, in physical units the
critical delay is $\tau_c=\pi\gamma/2U''(0)$, so stiffer minimum
destabilizes at shorter delays.
Past the instability the local dynamics are the same for every smooth
well: near the minimum, every smooth potential with a well-defined
curvature is harmonic. The nonlinearity governs only the later,
large-amplitude motion. When the growing oscillation runs unimpeded to
the boundary, a single amplification event carries the particle out.
This is the case for the harmonic trap and for hardening wells, whose
curvature $U''$ grows with displacement. We call this the {\it slingshot} and develop it in
what follows.


\emph{The slingshot law.}---In the slingshot regime the escape unfolds in two stages. During the
first delay interval the particle diffuses freely, accumulating a
small Gaussian displacement $\sim\sqrt{2D\tau}$; for a destabilized minimum, this displacement seeds a growing mode that is amplified deterministically to the boundary, the exponential growth outpacing further noise. The two stages separate cleanly for this reason, and the escape time is set not by waiting for a rare fluctuation but by the time this deterministic growth needs to magnify a typical seed, which depends only logarithmically on how small the seed is. The linearized
trajectory past threshold is therefore
\begin{equation}
x(t)=R\,e^{\mu t}\cos(\omega t-\varphi),\qquad
\mu+i\omega=\lambda=z/\Trel,
\label{eq:traj}
\end{equation}
with growth rate $\mu$ and frequency $\omega$ set by the leading unstable roots 
$z$ of Eq.~\eqref{eq:char}. The amplitude $R$ and phase $\varphi$ are the only stochastic quantities: they are fixed by the noise of the seeding stage and then freeze, since noise arriving later is amplified
for less time and is negligible against the growing solution \cite{SM}. The escape is thus controlled entirely by the statistics of this seed: $R$
is the effective random initial displacement handed to the instability,
and because it is set by a brief window of Gaussian noise it is itself
Gaussian-distributed \cite{SM}.

Escape occurs when for the first time the particle arrives at $\pm L$ so that one should have \
$R\,e^{\mu\mathcal{T}}|\cos(\omega\mathcal{T}-\varphi)|=L$, and thus upon rewriting
\begin{equation}
\mu\,\mathcal{T}
=\underbrace{\ln\frac{L}{R}}_{\text{envelope}}
\;-\;\underbrace{\ln\big|\cos(\omega\mathcal{T}-\varphi)\big|}_{\text{oscillation}} .
\label{eq:split}
\end{equation}
The envelope term needs only a Gaussian average $\langle\ln R\rangle$ and the oscillation term is bounded and so cannot
change the leading barrier dependence. 
Carrying out both averages, where
$L^2=2\Delta U\tau_{\rm relax}/\gamma$ relates the boundary to the
barrier, and writing the result per relaxation time, gives the escape
law below (see End Matter for full derivation)
\begin{equation}
\frac{\langle\mathcal{T}\rangle}{\Trel}
=A(s)\,\ln\frac{\Delta U}{D}+B(s),
\label{eq:law}
\end{equation}
which is the central result of this work. 
Every quantity follows from the one complex root $z$. The slope comes
entirely from the envelope term, through the growth rate
$\mu=\mathrm{Re}\,z/\Trel$:
\begin{equation}
A(s)=\frac{1}{2\,\mathrm{Re}\,z},
\label{eq:A}
\end{equation}
and $B(s)$ is the intercept (worked out at End Matter).

\begin{figure*}[t]
\centering
\includegraphics[width=\textwidth]{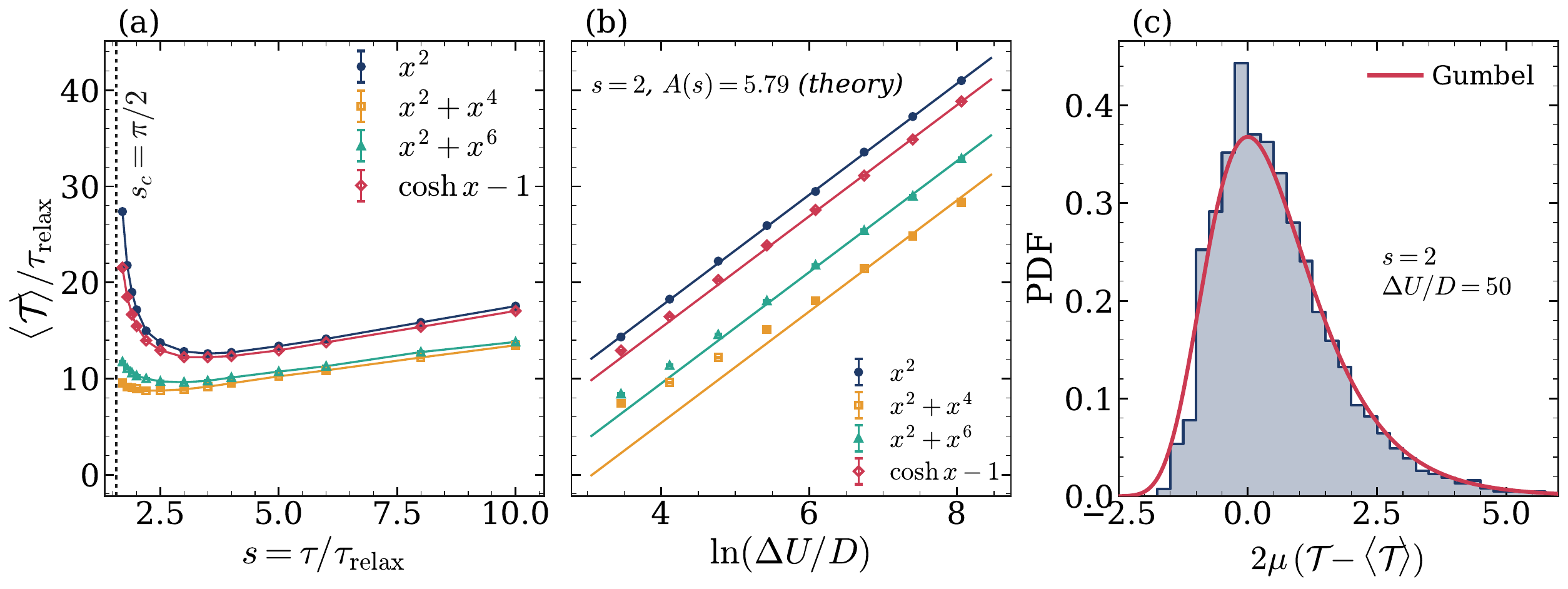}
\caption{
\textbf{Left: Non-monotone speedup.} Dimensionless mean escape time
versus delay at $\Delta U/D=50$, for the four potentials. Each falls sharply past $s_c=\pi/2$ to a
minimum near $s\simeq3$ --- exponentially faster than Arrhenius, and
below the free-diffusion time --- then saturates to the diffusive
limit in large $s$. 
The non-monotonic shape is a generic feature of slingshot escape. 
\textbf{Middle: Universality of the logarithmic slope.}  Dimensionless mean escape time versus $\ln(\Delta U/D)$ at $s=2$,
for four hardening potentials 
(harmonic, $x^2+x^4$, $x^2+x^6$, $\cosh x -1$). Lines share the single
predicted slope $A(s)$ with one fitted intercept each. 
\textbf{Right: Escape-time distribution is Gumbel.} Histogram of the
rescaled, centered escape time
$2\mu(\mathcal{T}-\langle\mathcal{T}\rangle)$ from Langevin simulation
at $s=2$ and the accessible barrier $\Delta U/D=50$ for the harmonic
potential, against the standard Gumbel density (solid). The
positive skew reflects long escapes seeded by small
fluctuations. The numerical code takes $\gamma=L=1$ throughout.  }
\label{fig:universal}
\end{figure*}

Eq.~\eqref{eq:law} states that Arrhenius exponential $e^{\Delta U/D}$
is replaced by a logarithm, whose coefficient $A(s)$ is a pure function
of the delay. 
The acceleration is non-monotone in $s$: it drives
$\langle\mathcal{T}\rangle$ below even the free-diffusion time $\tau_D=L^2/D$
near an optimal delay $s\simeq3$, so the particle leaves faster than it
would with no trap at all. The 
non-monotone shape is generic and appears for every hardening potential (Fig.~\ref{fig:universal}a). We now show that the slope $A(s)$ of
Eq.~\eqref{eq:law} is a curvature invariant, identical for every
potential in this class, while the intercept $B(s)$ alone reflects the
well's shape. 

\emph{Universality of the slope.}---  In the weak-noise
regime $\Delta U/D\gg1$ the seed handed to the instability is small,
$R\sim\sqrt{2D\tau}$, so the particle begins its climb deep inside the
harmonic neighborhood of the minimum and is amplified there at the
rate $\mu=\mathrm{Re}\,z/\tau_{\rm relax}$ fixed by the curvature
$U''(0)$ alone. For hardening potentials, the boundary sits at $L\propto\sqrt{\Delta U}$
while $R\propto\sqrt{D}$, implies 
$\ln(L/R)\simeq\tfrac12\ln(\Delta U/D)$, with the crossing
time $\mu^{-1}\ln(L/R)$ delivers the universal slope $A(s)$. That anharmonic shape of the potential is felt only in the final approach, and is associated with the lag due to oscillations. The latter contributes to $B(s)$ (see End Matter).

Fig.~\ref{fig:universal}b confirms this. For four hardening potentials, the escape time is linear in $\ln(\Delta U/D)$ with
a common slope $A(s)$ and four distinct intercepts; the residuals to the
parameter-free slope vanish as $\Delta U/D$ grows. The wells differ in
stiffness by a large factor at the boundary, yet their asymptotic slopes
coincide: escape from the delay-induced instability probes the curvature
of the minimum, not the shape of the barrier.

That curvature-only dependence is itself a specific instance of a
general fact: a logarithmic-in-barrier escape time is the generic
signature of noise-seeded escape near a linearly unstable point. There,
amplification by an instability of rate $\lambda_c$ gives an escape
time $\lambda_c^{-1}\ln(\Delta U/D)$ that simply counts the separation
between seed and boundary. The same structure underlies the decay of a
prepared unstable state \cite{suzuki1977,haake1978} and the mean transition-path time
across a barrier, $\tau_{\rm TP}\sim\lambda_b^{-1}\ln(\Delta U/D)$ with
$\lambda_b$ the saddle curvature \cite{chung2009,chaudhury2010harmonic,neupane2016direct,laleman2017} --- but in both cases
$\lambda_c$ or $\lambda_b$ is whatever the prepared state or the given
well provides, with nothing to tune. Here, by contrast, the same well
is carried from stable confinement through the Hopf threshold and
along the entire $A(s)$ family by varying $s=\tau/\tau_{\rm relax}$
alone (Fig.~\ref{fig:universal}b): the delay knob, not the shape of the potential,
decides whether the escape is logarithmic, and locates the slingshot
within this family as the case where the tunable rate is the curvature
of the minimum. Consequently, $A(s)$ depends on the potential only through $U''(0)$,
revealing a universal slope across the entire slingshot class,
independent of the well's shape.

\emph{Escape-time distribution.}---The theory predicts not only the mean
but the full distribution. The seed fixes both the amplitude $R$ and the
phase $\varphi$ of the growing mode in Eq.~\eqref{eq:traj}. Near threshold, where
$\mathrm{Re}\,z\ll|\mathrm{Im}\,z|$, the mode turns many times before it
grows appreciably, so the phase $\varphi$ is left uniform and independent
of $R$. The bounded oscillation term in Eq.~\eqref{eq:split} then decouples from the envelope and averages to a constant, and the escape statistics are
inherited from $\ln R$ alone (at large barriers; see \cite{SM}). Histograms of
$2\mu(\mathcal{T}-\langle\mathcal{T}\rangle)$ collapse onto the Gumbel
form, with the characteristic positive skew reflecting long escapes
seeded by small fluctuations, already at the moderate, experimentally
accessible barrier $\Delta U/D=50$ (Fig.~\ref{fig:universal}c). This holds near threshold
and up to moderate $s$; as $s$ grows the mode grows appreciably within a
single turn, the phase $\varphi$ ceases to be uniform and becomes
correlated with $R$, and the distribution departs from Gumbel \cite{gumbel1958,hoyt1947}.

It is moreover insensitive to
how the particle is prepared: any zero-mean Gaussian initial condition
--- a thermally equilibrated one in particular --- changes only the
seed variance, which cancels upon centering \cite{SM}. Because the seed
is generated in the linear region, where every smooth well is harmonic,
the Gumbel form is a property of the slingshot class rather than of the
harmonic trap; its convergence across potentials is shown in \cite{SM}, and in particular the universal collapse in Fig.~S3.

\emph{Discussion.}--- In this \textit{Letter}, we have shown that a time delayed feedback in a conservative
force removes the Arrhenius bottleneck of thermal escape. 
Beyond the threshold, the
well is destabilized and a typical thermal fluctuation is amplified deterministically. In this slingshot regime the amplification runs directly to the boundary and the mean escape time is logarithmic in the barrier height; we obtained this law in closed form, showed its slope to be a curvature invariant
shared across hardening potentials, and predicted the Gumbel escape-time
distribution, verifying each against detailed numerical simulations.

Delay-induced destabilization of a trapped particle is experimentally
established: a colloidal bead under continuous time-delayed optical
feedback~\cite{belldavies2023} and a levitated nanoparticle with
feedback applied through a separate beam~\cite{debiossac2020,
debiossac2022} both lose stability at a critical, software-set delay
--- in both cases the same Hopf bifurcation studied here. Those experiments
retain an instantaneous trap alongside the delayed force, which competes
with it: for instantaneous and delayed stiffnesses $k_0,k_1$ the
threshold becomes
$s_c=\arccos(-r)/\sqrt{1-r^{2}}$ with $r=k_0/k_1$~\cite{hayes1950}, rising from $\pi/2$
at $r=0$ and diverging as $r\to1$. The regime treated here is thus
reached by making the delayed force dominant, $k_1>k_0$. For micron-scale colloids the relaxation time is $\tau_{\rm relax}=\gamma/\Delta U\sim100\,\mu$s--$10\,$ms, and the delays in this range are directly accessible \cite{belldavies2023,debiossac2020,debiossac2022},
placing the slingshot threshold $s_c=\pi/2$ within reach.

The purpose of this work was to highlight an effective delay-induced escape mechanism namely slingshot which is shared by a broad class of potentials. 
There could be a
 second escape route that appears when the growing oscillation does not run
to the boundary but saturates to a limit cycle. For a softening well the restoring force
weakens with displacement, and the instability settles onto a stable
limit cycle of finite amplitude below the boundary; escape then requires
a further thermal fluctuation over the residual barrier, an Arrhenius
law with a delay-reduced effective barrier. This route
is not logarithmic, and its quantitative theory is developed elsewhere;
delayed nonlinear dynamics can still produce more complex trajectories
\cite{wernecke2019}, which we do not classify here. Both routes,
however, share the exponential speedup, so the acceleration is a generic
consequence of the delay-induced instability.

This last point suggests a use for the effect. Because the threshold and the growth rate are fixed by the curvature $U''(0)$, scanning the delay and recording escape statistics measures the curvature at the bottom of a well without ever crossing its barrier at the Arrhenius rate. Locating minima and characterizing their basins is a standing problem in the sampling of rough free-energy landscapes, addressed by biasing and resetting schemes in statistical physics \cite{laio2002,blumer2022,Evans2011Diffusion,Evans2020Stochastic,pal2024random}. A delayed force attacks it from a different direction, by removing the metastability rather than reweighting it.   In higher dimensions the instability appears first along the stiffest Hessian direction, so the delay should select escape channels by curvature rather than barrier height; mapping this out, together with the underdamped case and delayed forces in interacting systems and active matter \cite{Holubec2021,kopp2023,Wang2023}, are natural next steps.

\textit{Acknowledgments.--}
We thank Yael Roichman, Tomer Markovich, Shlomi Reuveni and Timo Schorlepp for useful discussions. 
A.P. acknowledges research funding under the
scheme ANRF/ARGM/2025/001623 from ANRF, India and research support from the
Department of Science and Technology, India, SERB Start-up Research Grant Number
SRG/2022/000080. A.P. also acknowledges the International Research Project (IRP)
titled “Classical and quantum dynamics in out of equilibrium systems” by CNRS,
France. V.K. and A.P. gratefully acknowledge research support from the Department of
Atomic Energy, Government of India via Soft Matter Apex projects.

\bibliography{references}

@article{arrhenius1889,
  author  = {S. Arrhenius},
  title   = {},
  journal = {Zeitschrift für Physikalische Chemie},
  volume  = {4},
  pages   = {226},
  year    = {1889}
}

@article{kramers1940,
  author  = {H. A. Kramers},
  title   = {},
  journal = {Physica},
  volume  = {7},
  pages   = {284--304},
  year    = {1940}
}

@incollection{pal2024random,
  title={Random resetting in search problems},
  author={Pal, Arnab and Stojkoski, Viktor and Sandev, Trifce},
  booktitle={Target Search Problems},
  pages={323--355},
  year={2024},
  publisher={Springer}
}

@article{kopp2026inference,
  title={Inference of a time delay in stochastic systems},
  author={Kopp, Robin A and Klapp, Sabine HL and Gupta, Deepak},
  journal={Physical Review Research},
  volume={8},
  number={1},
  pages={013129},
  year={2026},
  publisher={APS}
}

@article{geiss2019brownian,
  title={Brownian molecules formed by delayed harmonic interactions},
  author={Geiss, Daniel and Kroy, Klaus and Holubec, Viktor},
  journal={New Journal of Physics},
  volume={21},
  number={9},
  pages={093014},
  year={2019},
  publisher={IOP Publishing}
}

@book{yi2010time,
  title={Time-delay systems: {A}nalysis and control using the {L}ambert {W} function},
  author={Yi, Sun and Nelson, Patrick W. and Ulsoy, A. Galip},
  year={2010},
  publisher={World Scientific},
  address={Singapore}
}

@article{hanggi1990,
  author  = {P. H{\"a}nggi and P. Talkner and M. Borkovec},
  title   = {},
  journal = {Reviews of Modern Physics},
  volume  = {62},
  pages   = {251--341},
  year    = {1990}
}

@article{grote1980,
  author  = {R. F. Grote and J. T. Hynes},
  title   = {},
  journal = {Journal of Chemical Physics},
  volume  = {73},
  pages   = {2715},
  year    = {1980}
}

@article{kappler2018,
  author  = {J. Kappler and J. O. Daldrop and F. N. Br{\"u}nig and M. D. Boehle and R. R. Netz},
  title   = {},
  journal = {Journal of Chemical Physics},
  volume  = {148},
  pages   = {014903},
  year    = {2018}
}

@article{woillez2019,
  author  = {E. Woillez and Y. Zhao and Y. Kafri and V. Lecomte and J. Tailleur},
  title   = {},
  journal = {Physical Review Letters},
  volume  = {122},
  pages   = {258001},
  year    = {2019}
}

@article{kumar2024arrhenius,
  title={Arrhenius law for interacting diffusive systems},
  author={Kumar, Vishwajeet and Pal, Arnab and Shpielberg, Ohad},
  journal={Physical Review E},
  volume={109},
  number={3},
  pages={L032101},
  year={2024},
  publisher={APS}
}

@article{kumar2025inferring,
  title={Inferring intermediate states by leveraging the many-body Arrhenius law},
  author={Kumar, Vishwajeet and Pal, Arnab and Shpielberg, Ohad},
  journal={The Journal of Chemical Physics},
  volume={163},
  number={22},
  year={2025},
  publisher={AIP Publishing}
}

@article{kumar2024,
  author  = {V. Kumar and A. Pal and O. Shpielberg},
  title   = {},
  journal = {Journal of Chemical Physics},
  volume  = {160},
  pages   = {134107},
  year    = {2024}
}

@article{langer1969,
  author  = {J. S. Langer},
  title   = {},
  journal = {Annals of Physics},
  volume  = {54},
  pages   = {258},
  year    = {1969}
}

@article{bratsun2005,
  author  = {D. Bratsun and D. Volfson and L. S. Tsimring and J. Hasty},
  title   = {},
  journal = {Proceedings of the National Academy of Sciences},
  volume  = {102},
  pages   = {14593},
  year    = {2005}
}

@book{loos2021book,
  author    = {S. A. M. Loos},
  title     = {Stochastic Systems with Time Delay},
  publisher = {Springer},
  address   = {Cham},
  year      = {2021}
}

@article{holubec2021,
  author  = {V. Holubec and K. Geiss and S. A. M. Loos and K. Kroy and F. Cichos},
  title   = {},
  journal = {Physical Review Letters},
  volume  = {127},
  pages   = {258001},
  year    = {2021}
}

@article{berner2018,
  author  = {J. Berner and B. M{\"u}ller and J. R. Gomez-Solano and M. Kr{\"u}ger and C. Bechinger},
  title   = {},
  journal = {Nature Communications},
  volume  = {9},
  pages   = {999},
  year    = {2018}
}

@article{militaru2021,
  author  = {A. Militaru and others},
  title   = {},
  journal = {Nature Communications},
  volume  = {12},
  pages   = {2446},
  year    = {2021}
}

@article{suzuki1977,
  author  = {M. Suzuki},
  title   = {},
  journal = {Journal of Statistical Physics},
  volume  = {16},
  pages   = {11},
  year    = {1977}
}

@article{haake1978,
  author  = {F. Haake},
  title   = {},
  journal = {Physical Review Letters},
  volume  = {41},
  pages   = {1685},
  year    = {1978}
}

@article{mijalkov2016,
  author  = {M. Mijalkov and A. McDaniel and J. Wehr and G. Volpe},
  title   = {},
  journal = {Physical Review X},
  volume  = {6},
  pages   = {011008},
  year    = {2016}
}

@article{tsimring2001,
  title={Noise-induced dynamics in bistable systems with delay},
  author={Tsimring, Lev S and Pikovsky, Arkady},
  journal={Physical Review Letters},
  volume={87},
  number={25},
  pages={250602},
  year={2001},
  publisher={APS}
}

@article{pyragas1992,
  title={Continuous control of chaos by self-controlling feedback},
  author={Pyragas, Kestutis},
  journal={Physics letters A},
  volume={170},
  number={6},
  pages={421--428},
  year={1992},
  publisher={Elsevier}
}

@article{janson2004,
  author  = {N. B. Janson and A. G. Balanov and E. Sch{\"o}ll},
  title   = {},
  journal = {Physical Review Letters},
  volume  = {93},
  pages   = {010601},
  year    = {2004}
}

@article{guillouzic1999,
  author  = {S. Guillouzic and I. L'Heureux and A. Longtin},
  title   = {},
  journal = {Physical Review E},
  volume  = {59},
  pages   = {3970},
  year    = {1999}
}

@article{guillouzic2000,
  author  = {S. Guillouzic and I. L'Heureux and A. Longtin},
  title   = {},
  journal = {Physical Review E},
  volume  = {61},
  pages   = {4906},
  year    = {2000}
}

@book{zwanzig2001,
  author    = {R. Zwanzig},
  title     = {Nonequilibrium Statistical Mechanics},
  publisher = {Oxford University Press},
  address   = {Oxford},
  year      = {2001}
}

@book{gumbel1958,
  author    = {E. J. Gumbel},
  title     = {Statistics of Extremes},
  publisher = {Columbia University Press},
  address   = {New York},
  year      = {1958}
}

@article{hoyt1947,
  author  = {R. S. Hoyt},
  title   = {},
  journal = {Bell System Technical Journal},
  volume  = {26},
  pages   = {318},
  year    = {1947}
}

@article{wernecke2019,
  author  = {H. Wernecke and B. S{\'a}ndor and C. Gros},
  title   = {},
  journal = {Physics Reports},
  volume  = {824},
  pages   = {1},
  year    = {2019}
}

@article{laio2002,
  author  = {A. Laio and M. Parrinello},
  title   = {},
  journal = {Proceedings of the National Academy of Sciences},
  volume  = {99},
  pages   = {12562},
  year    = {2002}
}

@article{blumer2022,
  author  = {O. Blumer and S. Reuveni and B. Hirshberg},
  title   = {},
  journal = {Journal of Physical Chemistry Letters},
  volume  = {13},
  pages   = {11230},
  year    = {2022}
}

@article{hayes1950,
  author  = {N. D. Hayes},
  title   = {},
  journal = {Journal of the London Mathematical Society},
  volume  = {s1-25},
  pages   = {226},
  year    = {1950}
}

@article{Masoller2003,
  title     = {Distribution of Residence Times of Time-Delayed Bistable Systems Driven by Noise},
  author    = {Masoller, C.},
  journal   = {Physical Review Letters},
  volume    = {90},
  issue     = {2},
  pages     = {020601},
  numpages  = {4},
  year      = {2003},
  month     = {Jan},
  publisher = {American Physical Society},
  doi       = {10.1103/PhysRevLett.90.020601},
  url       = {https://link.aps.org/doi/10.1103/PhysRevLett.90.020601}
}

@article{chaudhury2010harmonic,
  author    = {Chaudhury, Srabanti and Makarov, Dmitrii E.},
  title     = {A harmonic transition state approximation for the duration of reactive events in complex molecular rearrangements},
  journal   = {The Journal of Chemical Physics},
  volume    = {133},
  number    = {3},
  pages     = {034118},
  year      = {2010},
  publisher = {AIP Publishing},
  doi       = {10.1063/1.3459058},
  url       = {https://doi.org}
}

@article{Wang2023,
  title = {Spontaneous vortex formation by microswimmers with retarded attractions},
  author = {Wang, Xiangzun and Chen, Pin-Chuan and Kroy, Klaus and Holubec, Viktor and Cichos, Frank},
  journal = {Nature Communications},
  volume = {14},
  number = {1},
  pages = {56},
  year = {2023},
  month = {Jan},
  publisher = {Nature Publishing Group UK London},
  doi = {10.1038/s41467-022-35427-7},
  url = {https://doi.org/10.1038/s41467-022-35427-7}
}

@article{Kopp2023,
  title = {Persistent motion of a Brownian particle subject to repulsive feedback with time delay},
  author = {Kopp, Robin A. and Klapp, Sabine H. L.},
  journal = {Physical Review E},
  volume = {107},
  issue = {2},
  pages = {024611},
  numpages = {12},
  year = {2023},
  month = {Feb},
  publisher = {American Physical Society},
  doi = {10.1103/PhysRevE.107.024611},
  url = {https://link.aps.org/doi/10.1103/PhysRevE.107.024611}
}

@article{Evans2011Diffusion,
  title     = {Diffusion with Stochastic Resetting},
  author    = {Evans, Martin R. and Majumdar, Satya N.},
  journal   = {Physical Review Letters},
  volume    = {106},
  issue     = {16},
  pages     = {160601},
  numpages  = {4},
  year      = {2011},
  month     = {Apr},
  publisher = {American Physical Society},
  doi       = {10.1103/PhysRevLett.106.160601},
  url       = {https://link.aps.org/doi/10.1103/PhysRevLett.106.160601}
}

@article{Evans2020Stochastic,
  title     = {Stochastic resetting and applications},
  author    = {Evans, Martin R. and Majumdar, Satya N. and Schehr, Gr{\'e}gory},
  journal   = {Journal of Physics A: Mathematical and Theoretical},
  volume    = {53},
  number    = {19},
  pages     = {193001},
  year      = {2020},
  month     = {Apr},
  publisher = {IOP Publishing},
  doi       = {10.1088/1751-8121/ab7cfe},
  url       = {https://doi.org/10.1088/1751-8121/ab7cfe}
}

@article{neupane2016direct,
  author    = {Neupane, Krishna and Foster, David A. N. and Dee, Derek R. and Yu, Hao and Wang, Feng and Woodside, Michael T.},
  title     = {Direct observation of transition paths during the folding of proteins and nucleic acids},
  journal   = {Science},
  volume    = {352},
  number    = {6282},
  pages     = {239--242},
  year      = {2016},
  publisher = {American Association for the Advancement of Science},
  doi       = {10.1126/science.aad0637},
  url       = {https://doi.org}
}

@article{masoller2002noise,
  title     = {Noise-Induced Resonance in Delayed Feedback Systems},
  author    = {Masoller, C.},
  journal   = {Physical Review Letters},
  volume    = {88},
  issue     = {3},
  pages     = {034102},
  numpages  = {4},
  year      = {2002},
  month     = {Jan},
  publisher = {American Physical Society},
  doi       = {10.1103/PhysRevLett.88.034102},
  url       = {https://link.aps.org/doi/10.1103/PhysRevLett.88.034102}
}

@article{Curtin,
  title = {Distribution of residence times in bistable noisy systems with time-delayed feedback},
  author = {Curtin, D. and Hegarty, S. P. and Goulding, D. and Houlihan, J. and Busch, Th. and Masoller, C. and Huyet, G.},
  journal = {Physical Review E},
  volume = {70},
  issue = {3},
  pages = {031103},
  numpages = {4},
  year = {2004},
  month = {Sep},
  publisher = {American Physical Society},
  doi = {10.1103/PhysRevE.70.031103},
  url = {https://link.aps.org/doi/10.1103/PhysRevE.70.031103}
}

@article{Loos2017Force,
  title = {Force-linearization closure for non-Markovian Langevin systems with time delay},
  author = {Loos, Sarah A. M. and Klapp, Sabine H. L.},
  journal = {Physical Review E},
  volume = {96},
  issue = {1},
  pages = {012106},
  numpages = {11},
  year = {2017},
  month = {Jul},
  publisher = {American Physical Society},
  doi = {10.1103/PhysRevE.96.012106},
  url = {https://link.aps.org/doi/10.1103/PhysRevE.96.012106}
}

@article{Loos2019Heat,
  title = {Heat flow due to time-delayed feedback},
  author = {Loos, Sarah A. M. and Klapp, Sabine H. L.},
  journal = {Scientific Reports},
  volume = {9},
  number = {1},
  pages = {2491},
  year = {2019},
  month = {Feb},
  publisher = {Nature Publishing Group},
  doi = {10.1038/s41598-019-39320-0},
  url = {https://www.nature.com/articles/s41598-019-39320-0}
}

@article{Rosinberg2015Stochastic,
  title = {Stochastic thermodynamics of Langevin systems under time-delayed feedback control: Second-law-like inequalities},
  author = {Rosinberg, M. L. and Munakata, T. and Tarjus, G.},
  journal = {Physical Review E},
  volume = {91},
  issue = {4},
  pages = {042114},
  numpages = {15},
  year = {2015},
  month = {Apr},
  publisher = {American Physical Society},
  doi = {10.1103/PhysRevE.91.042114},
  url = {https://link.aps.org/doi/10.1103/PhysRevE.91.042114}
}

@misc{SM,
  title = {See Supplemental Material for additional derivations, numerical methods, and supporting figures.}
}

@article{goldstein1969,
  author  = {Goldstein, Martin},
  title   = {Viscous Liquids and the Glass Transition: A Potential Energy Barrier Picture},
  journal = {J. Chem. Phys.},
  volume  = {51},
  number  = {9},
  pages   = {3728--3739},
  year    = {1969},
  doi     = {10.1063/1.1672587}
}

@article{chung2009,
  author = {Chung, Hoi Sung and Louis, John M. and Eaton, William A.},
  title = {Experimental determination of upper bound for transition path times in protein folding from single-molecule photon-by-photon trajectories},
  journal = {Proc. Natl. Acad. Sci. USA},
  volume = {106},
  pages = {11837},
  year = {2009}
}

@article{laleman2017,
  author = {Laleman, M. and Carlon, E. and Orland, H.},
  title = {Transition path time distributions},
  journal = {J. Chem. Phys.},
  volume = {147},
  pages = {214103},
  year = {2017}
}

@article{belldavies2023,
  author = {Bell-Davies, M. and Curran, A. and Liu, Y. and Dullens, R. P. A.},
  title = {Dynamics of a colloidal particle driven by continuous time-delayed feedback},
  journal = {Phys. Rev. E},
  volume = {107},
  pages = {064601},
  year = {2023}
}

@article{debiossac2020,
  author = {Debiossac, M. and Grass, D. and Alonso, J. J. and Lutz, E. and Kiesel, N.},
  title = {Thermodynamics of continuous non-Markovian feedback control},
  journal = {Nat. Commun.},
  volume = {11},
  pages = {1360},
  year = {2020}
}

@article{debiossac2022,
  author = {Debiossac, M. and Rosinberg, M. L. and Lutz, E. and Kiesel, N.},
  title = {Non-Markovian feedback control and acausality: An experimental study},
  journal = {Phys. Rev. Lett.},
  volume = {128},
  pages = {200601},
  year = {2022}
}

\onecolumngrid
\vspace{1em}
\begin{center}\textbf{END MATTER}\end{center}
\vspace{0.5em}
\twocolumngrid

\noindent
\emph{Derivation of the logarithmic law:}
This section carries out in detail the averaging that turns the first-passage
condition Eq.~\eqref{eq:split} into the escape law Eq.~\eqref{eq:law}, and
gives the exact intercept $B(s)$ for the harmonic trap. The slope
$A(s)$ follows from the dominant time scale.

The difficulty in Eq.~\eqref{eq:split} is that $\mathcal{T}$ appears
implicitly on both sides, requiring a self-consistent solution.
Averaging Eq.~\eqref{eq:split}
directly is obstructed by its last term, $\ln|\cos(\omega\mathcal
T-\varphi)|$, evaluated at the escape time $\mathcal T$ itself. We resolve this by measuring time from the instant the deterministic
envelope $Re^{\mu t}$ reaches the boundary, i.e. making the split $\mathcal{T} = \mathcal{T}_{\rm env}+\Delta $. This defines the two
stages of the escape,
\begin{equation}
\mathcal T_{\rm env}=\frac1\mu\ln\frac{L}{R},\qquad
e^{\beta u}\,|\cos(u-a)|=1,     
\label{eq:env}
\end{equation}
where $a = \varphi - \omega \mathcal{T}_{\rm env}$, $\beta=\mu/\omega$ and $u=\omega\Delta$ is the rescaled phase. Hence, the escape time comprises the envelope-crossing time
$\mathcal T_{\rm env}$, determined by $R e^{\mu\mathcal T_{\rm env}}=L$,
and a residual lag $\Delta$ required for
$|\cos(\omega t-\varphi)|$ to reach unity.

Thus, in $\mathcal T_{\rm env}$ the self-reference is resolved: it is defined by
the envelope alone, and depends on the seed only through the amplitude
$R$, which is Gaussian with variance set by the noise $\propto
D$ (\cite{SM}, \S S6). Its average is therefore a single Gaussian integral.

Recalling $L^2=2\Delta U\tau_{\rm
relax}/\gamma$, and performing the average yields 
\begin{equation}
\label{eq:av env}
\frac{\langle\mathcal T_{\rm env}\rangle}{\tau_{\rm relax}}
=A(s)\ln\frac{\Delta U}{D}+{\rm const},\qquad A(s)=\frac{1}{2\,{\rm
Re}\,z},
\end{equation}
which is the leading part of \eqref{eq:law}: the slope comes entirely
from the envelope.

The lag term $\Delta$ therefore adds as the remaining constant.  The envelope crossing carries the whole barrier dependence, growing as $\tfrac12
\ln(\Delta U/D)$ and diverging at weak noise, whereas $\Delta$ is at
most a quarter period of the oscillation --- a bounded number of
relaxation times, independent of $\Delta U/D$. At the large $\Delta U/D$ limit and  where the slingshot law holds, $\mathcal
T_{\rm env}\gg\Delta$, so the envelope crossing alone sets the escape
time. The self-reference that obstructed Eq.~\eqref{eq:split} survives only inside
$\Delta$, where it is confined to the subleading correction and
cannot affect the slope.

We now evaluate that subleading term from Eq.~\eqref{eq:env}, by noticing that $a\in[0,\pi)$ is the phase from the envelope crossing to the next
cosine extremum, fixed by where in its cycle the oscillation sits when
the envelope arrives. Because $\mathcal T_{\rm env}$ depends
logarithmically on the Gaussian seed while the phase advances linearly,
a broad spread in $R$ winds the phase many times over when $\beta\ll1$, so $a$ is effectively uniform on $[0,\pi)$. The small $\beta$ limit is justified as $s\rightarrow s_c $ (Fig.~\ref{fig:roots}). Nevertheless, even at moderate $s>s_c $ value, where $\beta$ is not strictly negligible, the approximation still yields good results (also see \cite{SM}, \S S9).   

Thus, for a uniform $a$, the mean lag is
then a universal function of $\beta$ alone
\begin{equation}
\langle\Delta\rangle=\frac{\mathcal W(\beta)}{\omega},
\qquad \mathcal W(\beta)=\frac1\pi\int_0^\pi u(a;\beta)\,da\ \le\ \frac\pi2,
\label{eq:crossing}
\end{equation}
with the quarter-period bound. Collecting the envelope's constant with this wait gives the intercept $B(s)$. Matching to Eq.\eqref{eq:av env}, $B(s)$ is exactly the sum of the
constant left over from averaging $\ln R$ and the mean lag
\begin{align}
B(s) = &\underbrace{A(s)\Big[\ln\!\Big(\tfrac{1}{2\,{\rm Re}\,z}
|1+sz|^2\Big)+\gamma_E-\ln\dfrac{|z|+|{\rm Im}\,z|}{2|z|}\Big]}
_{\text{const of \eqref{eq:av env}, from }\langle\ln R\rangle} \nonumber \\
&+\underbrace{\dfrac{\mathcal W({\rm Re}\,z/{\rm Im}\,z)}{{\rm
Im}\,z}}_{\langle\Delta\rangle/\tau_{\rm relax}},
\label{eq:B}
\end{align}
where $\gamma_E$ is the Euler constant. 
The first term is the constant deferred from \eqref{eq:av env} in
full \cite[\S S6]{SM}. The second term is
$\langle\Delta\rangle/\tau_{\rm relax}$ itself, the oscillation wait
derived above. Both are functions of $s, z(s)$ alone. Namely, they are independent of the barrier-to-noise $\Delta U/D$. This justifies the dominance of the slope  and the emergence of the Gumbel distribution at large $\Delta U/D$, as demonstrated in Fig.~\ref{fig:universal}(b,c).

We are now left with the evaluation of 
$\mathcal{W}$, which is non-trivial. To that end, we perform a small $\beta$ evaluation, at the crossing precedes the extremum at $u=a$. Consider then $u=a-\epsilon$, where both $\epsilon$ and $\beta$ are expanded to leading order. This leads to  $\epsilon^2 /2 + \beta (\epsilon-a)=0 $. So, for small $\beta$, we find $u(a;\beta)\simeq a-\sqrt{2\beta a}+\beta$. Plugging back this perturbative series into the expression of $\mathcal{W}$ in Eq. (\ref{eq:crossing}) and averaging over uniform
$a$ with $\langle a\rangle=\pi/2$ and $\langle\sqrt a\rangle=
\tfrac23\sqrt\pi$, we arrive at
\begin{equation}
    \mathcal W(\beta)=\frac\pi2-\frac23\sqrt{2\pi\beta}+\beta+O(\beta^{3/2}).
\end{equation}
The leading $\pi/2$ is the quarter-period bound: at $\beta=0$ the
envelope does not grow and the trajectory simply waits for the oscillatory term $|\cos|$ to
reach unity. For error estimation of this method, see \cite{SM} \S 9.2 for more details.

\end{document}